\documentclass[
 reprint,
 superscriptaddress,
 amsmath,amssymb,
 aps,
 prl
]{revtex4-2}

\usepackage{color}
\usepackage{amsmath}
\usepackage{pifont}
\usepackage{amssymb}  
\usepackage{bbold}
\usepackage{float}
\usepackage{placeins}
\usepackage{tikz}
\usepackage{makecell}
\usepackage{subfigure}
\usepackage{scalerel}
\usepackage{pifont}   
\usepackage{dcolumn}  
\usepackage{multirow} 
\usepackage{placeins}
\usepackage{soul}
\usepackage{xcolor}
\usepackage{graphicx}
\usepackage{dcolumn}
\usepackage{bm}
\usepackage{listings}

\begin{document}

\preprint{APS/123-QED}

\title {Doppler-free Rydberg spectroscopy in a warm vapor}

\author{Jeremy Glick}
\affiliation{DEVCOM Army Research Laboratory South, Austin, Texas 78712 USA
\looseness=-1}

\author{Brielle E. Anderson}
\affiliation{Department of Physics, The American University, Washington, DC, USA
\looseness=-1}

\author{T. Nathan Nunley}
\affiliation{General Technical Services, 1451 NJ-34, Wall Township, NJ 07727 USA 
\looseness=-1}
\affiliation{DEVCOM Army Research Laboratory South, Austin, Texas 78712 USA
\looseness=-1}

\author{Josiah Bingaman}
\affiliation{Department of Physics, Center for Complex Quantum Systems, The University of Texas at Austin, Austin, Texas 78712 USA  
\looseness=-1}

\author{Jian Jun Liu}
\affiliation{Department of Physics, Center for Complex Quantum Systems, The University of Texas at Austin, Austin, Texas 78712 USA  
\looseness=-1}

\author{David H. Meyer}
\affiliation{DEVCOM Army Research Laboratory, 2800 Powder Mill Rd, Adelphi MD 20783 USA
\looseness=-1}

\author{Paul Kunz}
\affiliation{DEVCOM Army Research Laboratory South, Austin, Texas 78712 USA
\looseness=-1}
\affiliation{Department of Physics, Center for Complex Quantum Systems, The University of Texas at Austin, Austin, Texas 78712 USA
\looseness=-1}

\date{\today}

\begin{abstract}

The common approach for producing Rydberg atoms in warm vapor cells is with lasers arranged in a counter-propagating, collinear configuration. Doppler effects in these configurations reduce the efficiency of excitation to the Rydberg state while also producing broadened spectral features. In this work, we demonstrate a three-laser Doppler-free excitation using laser beams whose k-vectors sum to zero, resulting in an enhancement in the Rydberg density and narrowed spectral features. A three-times enhancement to Rydberg density along with a near four-times reduction in spectroscopic line-widths are observed compared to a collinear configuration. This Doppler-free configuration could prove beneficial to Rydberg atomic technologies, such as electric field sensing with small volumes or deterministic photon sources.

\end{abstract}

\maketitle

\section{Introduction}

Atoms that have been excited to high principle quantum number, known as Rydberg atoms, are poised to be a leading platform for emerging quantum technologies. Having exaggerated properties, such as large polarizability and resonant dipole moments, Rydberg states enable high-fidelity qubit gate operations \cite{Endres_2025_GateFidelity, Evered-Lukin_2023_GateFidelity, Cao-Kauffman_2025_GateFidelity, Peper-Thompson_2025_GateFidelity} and quantum transduction \cite{Kumar-Simon_2023_Transduction}, quantum optical sources, and electromagnetic field sensors \cite{Adams-Shaffer_2020_Review}. 
Rydberg sensors utilizing warm vapor cells in particular may hold near-term promise, as they can serve as precise calibration tools \cite{Holloway_2017_selfCalibration, Anderson_2021_selfCalibration, Shaffer_2023_selfCalibration}, receive data over disparate frequency bands \cite{Meyer_2023_multiband, allinson2024simultaneous,meyer2018digital},  perform three-dimensional polarimetry \cite{elgee2024complete}, and reach the standard quantum limit and beyond \cite{Facon-Haroche_2016_RydbergCatSensor}. 

The most common method for probing Rydberg atoms for field sensing is using coherent spectroscopic techniques, such as electromagnetically induced transparency (EIT). These non-linear processes that coherently couple laser fields to the atomic states, can result in narrow spectral features  \cite{Marangos_2005_EIT, Raithel_2019_3photon,Finkelstein_2023_EIT}. Sensing of external radio-frequency (rf) electric fields then operates by observing spectroscopic shifts induced by the  Stark effect. Detection of very weak electric fields benefits from using many atoms to provide a strong signal, and having narrow spectral features such that small energy shifts can be resolved. 

While various mechanisms can broaden the coherent spectral feature, such as transit effects, power broadening, laser noise, atom-atom interactions, and stray background fields, Doppler effects consistently play an important role in warm vapor spectroscopy. They can  result in both broadening of the spectral features and reduced Rydberg excitation efficiency \cite{Su-Novikova_2025_RydbergNonColinear}. Typically two lasers are used in Rydberg spectroscopy, and this leaves residual Doppler shifts even when the two beams are counter-propagating, as their k-vectors do not fully cancel. Cesium atoms have a uniquely fortuitous excitation pathway that is nearly Doppler-free and collinear \cite{Shaffer_2023_3photon}, though one of the required lasers is currently a technological challenge.

In this work we demonstrate a three-photon ``star'' configuration whereby laser beams are angled to cancel residual Doppler shifts, which we call the Doppler-free (DF) configuration. While such star configurations have been investigated in the past both theoretically \cite{Ryabtsev_2011_starConfigThry} and experimentally \cite{grynberg1976, Sibalic-Weatherill_2016_ThreePhotSpinWave, guglielmo2025}, they typically have not achieved Rydberg spectral features narrower than the common two-laser resonances (i.e. approximately 10 MHz). We show a 1.18(8) MHz linewidth using a DF configuration, which is a four-times reduction compared to the counter-propagating collinear (CL) configuration with the same Rabi frequencies. Our DF linewidth is primarily limited by power broadening and Zeeman effects. Although the decreased overlap volume of the star configuration decreases the total integrated signal, we show there is a significant enhancement of Rydberg atom number densities for the star relative to the CL configurations. 

\begin{figure}[t]
    \centering
    \includegraphics[width=0.48\textwidth]{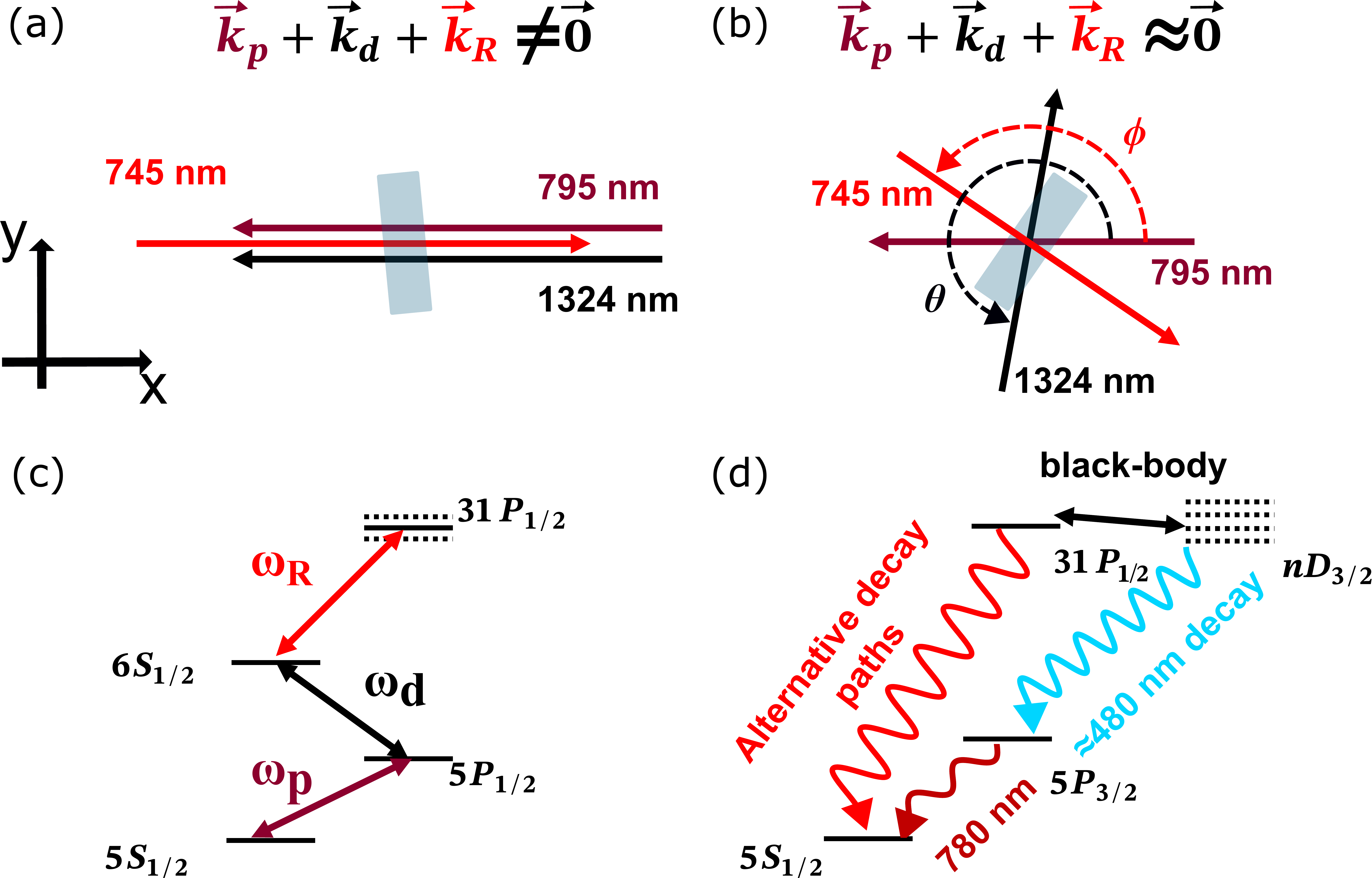}
    \caption{(a) CL configuration with all lasers having parallel linear polarization. (b) DF star configuration where $\theta$ and $\phi$ are the angles of the Rydberg and dressing lasers with respect to the probe. (c) Excitation level diagram used for both the DF and CL measurements. (d) Rydberg decay diagram for fluorescence measurements. We collect the $\approx$480 nm light.}
    \label{fig:system}
\end{figure}

\section{Motivation}

For Rydberg excitation within warm vapor cells, efficient coupling to the Rydberg state is hindered by residual Doppler shifts. In a three-laser excitation scheme with a probe, dressing, and Rydberg laser, the residual velocity dependent Doppler shift is 
\begin{equation}
    \delta(\Vec{v}) = \delta_0 + \left(\Vec{k_{\rm p}} + \Vec{k_{\rm d}}+\Vec{k_{\rm R}}\right)\cdot \Vec{v} \,.
\end{equation}
Here $\Vec{k_{\rm p}}$, $\Vec{k_{\rm d}}$, and $\Vec{k_{\rm R}}$ are the wave-vectors of the probe, dressing, and Rydberg lasers respectively. $\delta_0=\omega_{\rm ryd} - \omega_{\rm p} - \omega_{\rm d} - \omega_{\rm R}$ is the three-laser frequency detuning of the probe, dressing, and Rydberg laser frequencies $\omega_{\rm p}$, $\omega_{\rm d}$, and $\omega_{\rm R}$ from the excitation energy to the Rydberg state $\hbar \omega_{\rm ryd}$.

While three-laser schemes offer the ability to resonantly couple atomic energy levels and still cancel residual Doppler shifts, two-laser schemes are more restricted. Two-laser Doppler cancellation would require $|\Vec{k_{\rm p}}| = |\Vec{k_{\rm R}}|$, and to the best of our knowledge, there are no resonant atomic transitions that satisfy this criteria. With three-laser schemes, in addition to the fortuitous cesium CL excitation scheme \cite{Shaffer_2023_3photon}, there is the angular degree of freedom that enables complete cancellation of the residual excitation momentum, as depicted in Fig. \ref{fig:system}. Here the residual velocity-dependent Doppler shift can be expressed as,

\begin{eqnarray} \label{DFeqn}
    \delta\left(\Vec{v}\right) = \delta_0 + \left(k_{\rm p} + k_{\rm d} \cos\theta + k_{\rm R}\cos\phi \right) v_{\rm x} \nonumber\\ + \left(k_{\rm d}\sin\theta + k_{\rm R}\sin\phi\right) v_{\rm y} \,,
\end{eqnarray}
where $\theta$ and $\phi$ are the angles between the probe and dressing and the probe and Rydberg lasers respectively. For our excitation scheme, shown in Fig.~\ref{fig:system} (c), the residual Doppler shifts are canceled by setting $\theta \approx 4.526$ rad and $\phi \approx 2.556$ rad as measured from $\Vec{k_{\rm p}}$, depicted in Fig. \ref{fig:system} (b). 

\begin{figure}[t]
    \centering
    \includegraphics[width=0.45\textwidth]{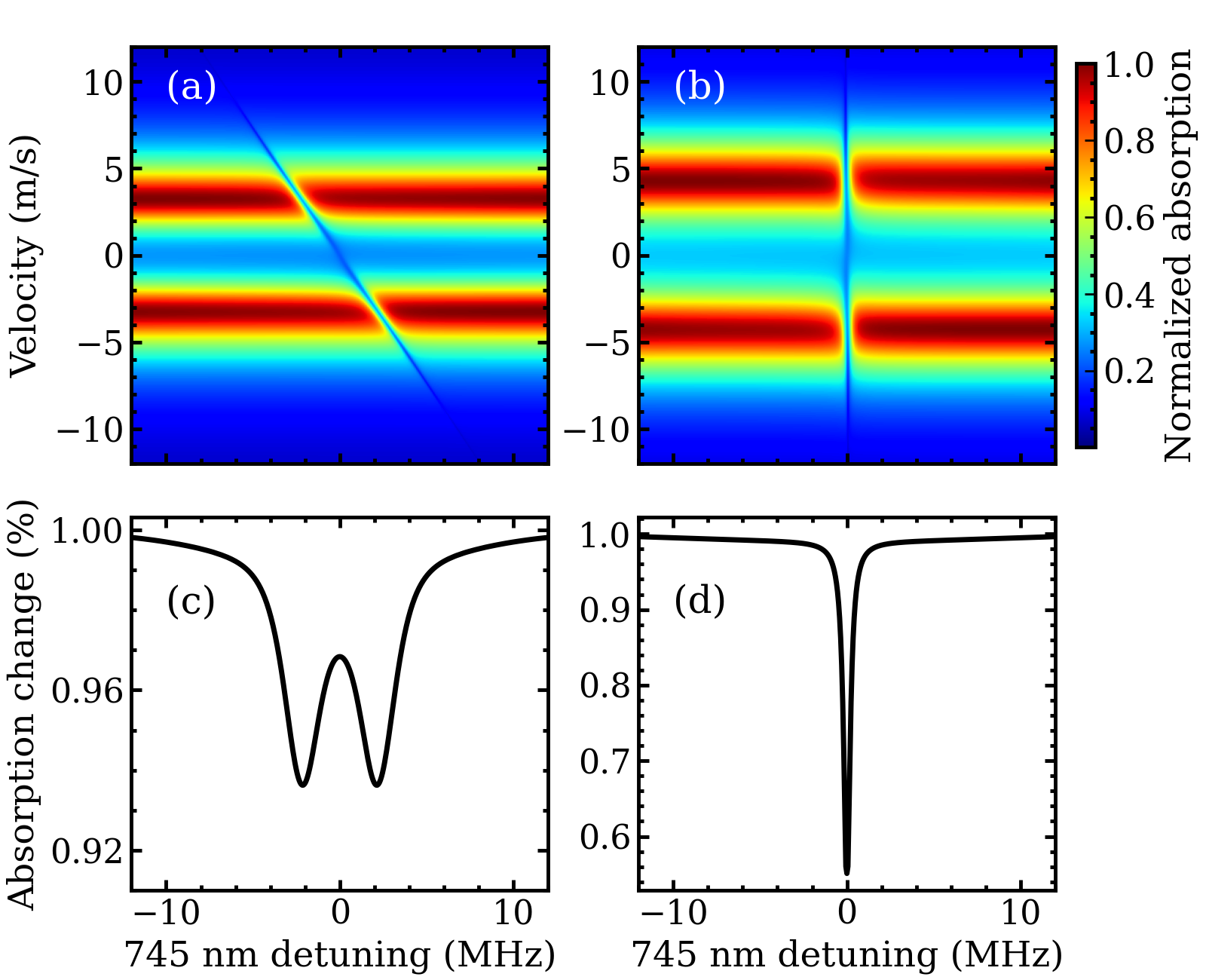}
    \caption{Top: One-dimensional simulation of the absorption of the probe laser per velocity class in a non-DF (a) and DF (b) configuration versus Rydberg laser detuning in a three-laser excitation scheme. Bottom: Simulated EIT spectra in the non-DF (c) and DF (d) configuration. Laser Rabi frequencies are $\Omega_{\rm p} = 2\pi\times 2$ MHz, $\Omega_{\rm d} = 2\pi\times10$ MHz, and $\Omega_{\rm R} = 2\pi\times 1$ MHz. }
    \label{fig:k_vec}
\end{figure}

 To illustrate the effect of residual k-vectors on the absorption of the probe laser, we simulate a non-DF and a DF three-laser scheme in one spatial dimension, as shown in Fig.~\ref{fig:k_vec}. The magnitude of the k-vectors in the non-DF simulation follow from the excitation scheme in Fig.~\ref{fig:system}(c) with the orientation following Fig.~\ref{fig:system}(a). In the DF simulation the magnitude of the k-vectors are $k_{\rm p} \approx 2\pi/795~{\rm nm}^{-1}$, $k_{\rm d} \approx -2\pi\cos\left(4.53\right)/1324 ~\rm{nm}^{-1}$ and $k_{\rm R}\approx-2\pi\cos\left(2.56\right)/745~\rm{nm}^{-1}$. Splitting of the EIT spectra is observed in the non-DF configuration compared to a single substantially narrower peak in the DF configuration. Further, there is a higher percentage of excitation in the DF configuration on resonance due to the cancellation of residual Doppler shifts. Notably, in this four-level scheme, simulations show that the largest contrast in absorption as a function of Rydberg laser detuning occurs for warm atoms with nonzero velocity classes rather than cold trapped atoms. Previous work based on three-level Lambda-schemes have shown similar counter-intuitive effects \cite{Javan-Scully_2002_DopplerEIT, Bashkansky_2005_DopplerEIT}. Our simulations here, and throughout the paper, are performed using the RydIQule python library \cite{Miller_2024_rydiqule, rydiqule_webpage} where the Linblad master equation is numerically solved in steady state for a 4-level system with Rabi frequencies $\Omega_{\rm p} = 2\pi\times2$ MHz, $\Omega_{\rm d} = 2\pi\times10$ MHz, and $\Omega_{\rm R}=2\pi\times1$ MHz. Decay rates between states are set to $\Gamma_{1\rightarrow0} = 2\pi\times5.7$ MHz, $\Gamma_{2\rightarrow1}=2\pi\times3.5$ MHz, and $\Gamma_{3\rightarrow2}=2\pi\times 0.01$ MHz. The Rabi frequencies and decay rates are motivated by experimental values.

\section{Method}
The experimental setup, shown in Fig.~\ref{fig:system}, utilizes a three-laser scheme to excite ground state $^{85}$Rb atoms to a Rydberg state within a cylindrical glass vapor cell, of 2.54 cm diameter and 5 mm length, containing rubidium in natural isotopic abundance. The vapor cell windows are heated to $\approx$100 $^{\circ}$C, while the stem is held at $\approx$60 $^{\circ}$C, to prevent rubidium condensation on the windows. For excitation to the Rydberg state, a probe laser at 795 nm excites the $5S_{1/2}(F=3)\rightarrow5P_{1/2}(F'=3)$ transition followed by the excitation from $5P_{1/2}(F=3)\rightarrow6S_{1/2}(F'=3)$ with a 1324 nm dressing laser and finally from $6S_{1/2}(F=3)\rightarrow 31P_{1/2}$ with a 745 nm Rydberg laser. Each laser has a linewidth of less than 200 kHz over all relevant timescales. The probe laser is frequency stabilized, via an offset phase-lock, to another laser locked to rubidium spectroscopy, and the dressing and Rydberg lasers are referenced to an ultra-stable optical cavity. At the Rydberg cell, the light is horizontally polarized and has Gaussian beam waists ($1/e^2$ value) of $\approx$1 mm. 

Spectroscopic EIT signals are observed by monitoring the transmission of the probe laser through the cell while scanning the frequency of the Rydberg laser. The signal is processed using a lock-in amplifier where amplitude modulation is performed on the Rydberg laser with an electro-optic amplitude modulator at modulation frequency of 122.915 kHz. To eliminate unwanted scattered 745 nm Rydberg laser light from reaching the detector, we spatially filter the transmitted probe with a single mode fiber and place a long-pass filter (cut-on wavelength of $\approx$750 nm) in front of the photo-diode detector. 

The alignment procedure for the laser beams is performed using a template with markings designating the positions of optical posts. For each laser, irises are mounted on the optical posts and are placed on the template before and after the cell. Each laser is aligned by maximizing the power through its respective irises. The positions of the apertures take into account refraction of the laser beams from the cell windows for our particular cell orientation. We estimate the uncertainty in our angular alignment procedure is $\pm3$ mrad for each laser beam. In the CL configuration, dichroic mirrors are used to overlap all three beams while preserving their relative polarizations with the alignment performed using a single pair of apertures. 

During the alignment procedure, it was found that slight changes to the angle of the vapor cell could result in a dramatic reduction of the EIT signal amplitude even when compensating for laser beam displacement. As the windows of our vapor cell do not have anti-reflective coatings, we are susceptible to reflections within the cell as well as possible etaloning effects. A full quantitative study of the effects from internal reflections is beyond the scope of the current work, but care was taken to position the angle of the cell such that the EIT amplitude was not compromised. 

Fluorescence measurements of the Rydberg atoms are performed to complement the transmission spectroscopy. A BlackFly BFS-PGE-70S7M-C CMOS camera \footnote{This is not an endorsement of this device or manufacturer, only a statement of its use in this case for completeness and benefit of the reader.} is positioned above the vapor cell, orthogonal to the plane of the laser beams. To differentiate between fluorescence of Rydberg states from that of the lower intermediate state, a band-pass filter is used to pass wavelengths of $\approx$462 to 512 nm. The blue fluorescence predominately occurs when atoms in the $31P_{1/2}$ state couple to an $nD_{3/2}$ state via blackbody radiation, which can then decay to a $5P$ state as shown in Fig.~\ref{fig:system}. Since we rely on imaging fluorescence from only a subset of Rydberg state decay channels, we do not use these images to directly quantify the absolute Rydberg number densities. Never-the-less, they allow for comparison of 
the relative densities between CL and DF configurations. Our fluorescence measurements operate with large probe and dressing Rabi frequencies to maximize signal, and this results in power broadening relative to the narrowest spectral features we observe when using small Rabi frequencies. To simplify the synchronization of the Rydberg laser scan and camera exposures, we do not stabilize the Rydberg laser frequency to our optical cavity during fluorescence imaging. 

\section{Results}

\begin{figure}[t]
    \centering
    \includegraphics[width=0.48\textwidth]{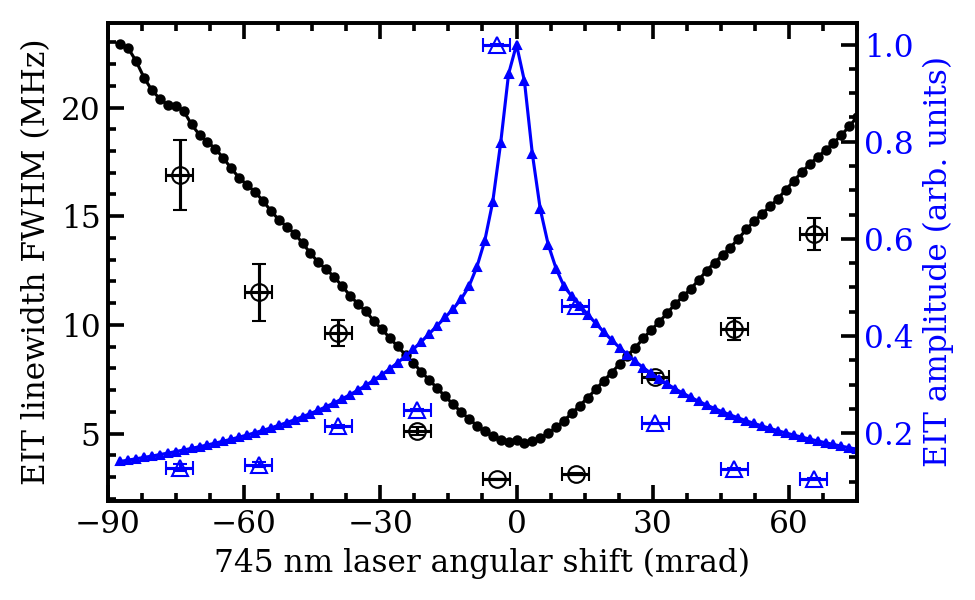}
    \caption{Amplitude and linewidth of EIT spectroscopy signal as the Rydberg-laser beam angle varies. Theory predicted values are solid closed symbols, and measured values are open symbols. The max signal amplitude and narrowest linewidth occur at the DF angle, denoted 0 mrad. For clarity of visual comparison of the profiles, a systematic offset in the experiment values has been compensated for by adding $13$ mrad. This should not be interpreted as absolute agreement of the optimum angular shift between theory and experiment.}
    \label{fig:angular_results}
\end{figure}

\begin{figure}[t!]
    \centering
    \includegraphics[width=0.48\textwidth]{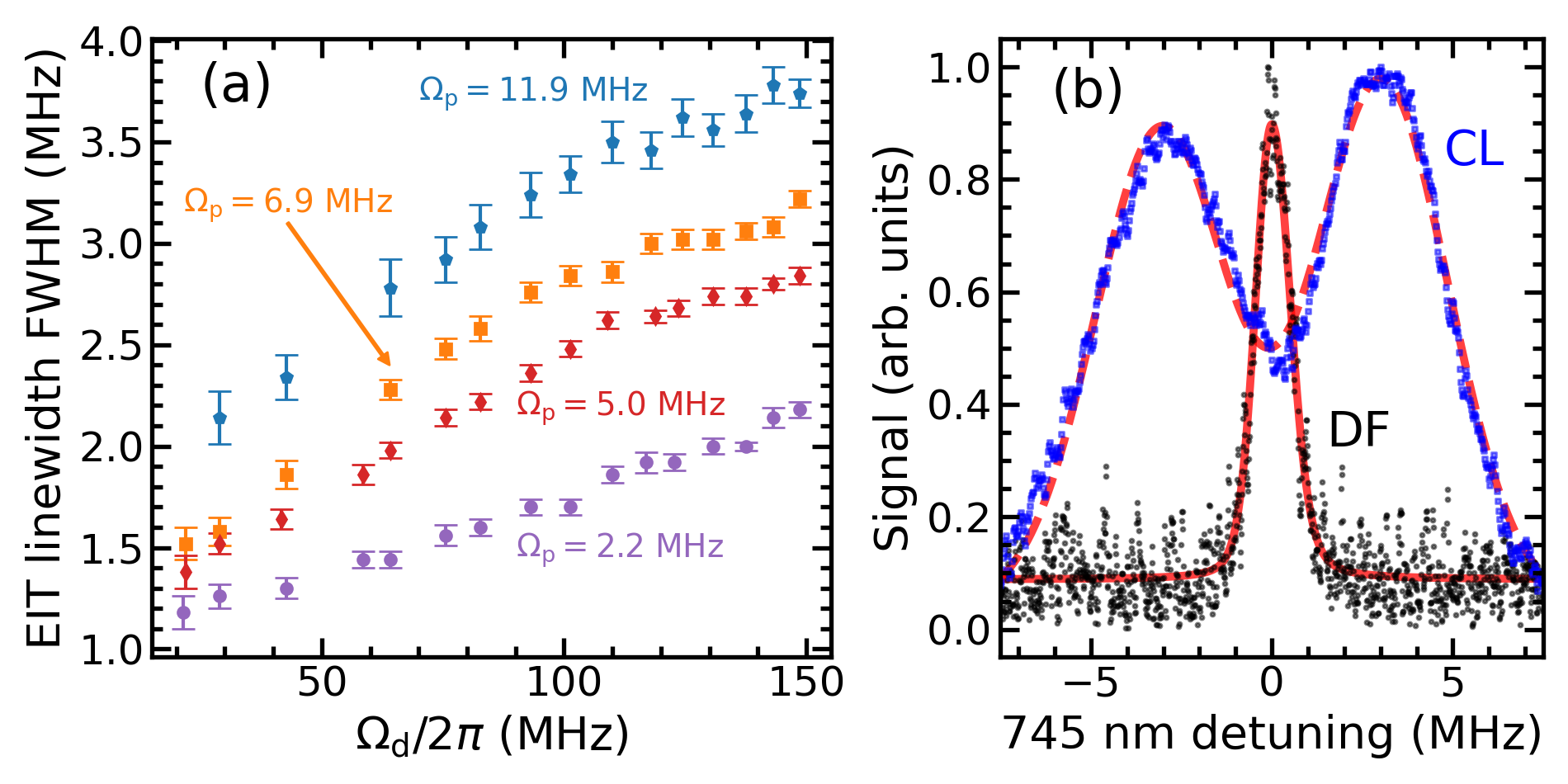}
    \caption{ (a) FWHM of the EIT linewidth in the DF configuration versus dressing laser Rabi frequency with Rydberg laser Rabi frequency held constant $\Omega_{R} = 1.2$ MHz and four probe laser Rabi frequencies represented by symbol color. (b) EIT spectra with the smallest linewidth from (a) showing FWHM of 1.18(8) MHz. Solid red curve is a Voigt fit. Comparable CL data is shown with blue squares along with a double Voigt fit (dashed red curve). For both CL and DF, the amplitudes have been scaled to a value of 1, and both use similar Rabi frequenices: DF probe Rabi frequency is $2\pi\times 2.2$ MHz and dressing Rabi frequency $2\pi \times21.4$ MHz, while the CL had $\Omega_{p} = 2\pi\times2.4$ MHz and $\Omega_{d} = 2\pi\times 20.5$ MHz } 
    \label{fig:power_broadening}
\end{figure}

To confirm that the laser configuration was at the optimal angle for DF excitation, EIT signals were collected over a range of Rydberg laser angles. These results are shown in Fig.~\ref{fig:angular_results} where 0$^\circ$ angular shift is the DF angle calculated from Equation \ref{DFeqn}. The values experimentally measured and those predicted (no free parameters) by the Lindblad master equation model are in rough agreement and show the same trends. We note there is a small residual systematic offset of  $\approx$13 mrad between theory and experiment, possibly due to effects of internal reflections from the cell as mentioned in Methods. For visual clarity of comparison, a positive $\approx$13 mrad shift has been added to the experimental results in Fig.~\ref{fig:angular_results}. A negative angular shift corresponds to a smaller angle between the Rydberg and probe lasers. The theoretical predictions include white phase-noise laser linewidths \cite{Plankensteiner_2016_LaserNoise, PritchardThesis_2012_LaserNoise} of 100 kHz, 100 kHz, and 10 kHz for the probe, dressing, and Rydberg lasers, respectively, but we do not include the modulation sidebands (which are approximately 122 kHz, and negligibly affect our results). The peak Rabi frequencies for these measurements are $\Omega_{p} = 2\pi\times 4.3$ MHz, $\Omega_{d} = 2\pi\times116.9$ MHz, and $\Omega_{R} = 1.2$ MHz with laser powers of $78\mu$W, 60mW, and 500 mW respectively. In order to have measurable EIT signal  at the larger angular shifts where the amplitude significantly falls, we use Rabi frequencies that are larger than those used for obtaining the narrowest linewidth shown in Fig.~\ref{fig:power_broadening}.

\begin{figure}[t!]
    \centering
    \includegraphics[width=0.46\textwidth]{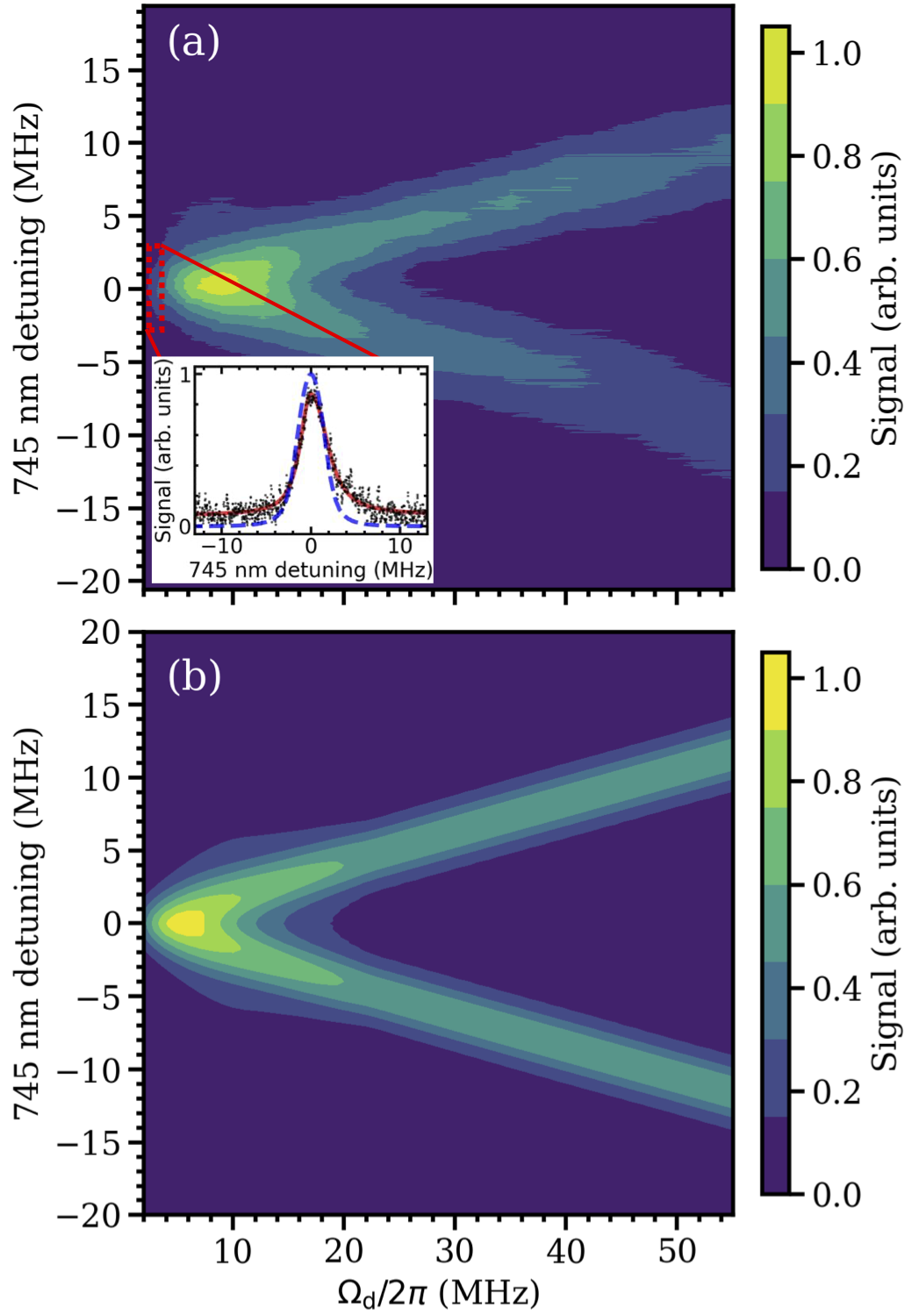}
    \caption{(a) Transmission spectra showing splitting and power broadening in the CL configuration as a function of Rydberg laser detuning and dressing laser Rabi frequency. The inset is a slice from the contour plot depicting experimental EIT spectra (black circles) at the lowest dressing Rabi frequency of $2\pi\times2$ MHz. The red solid line is a fit to the experimental data, and the blue dashed line is the simulated EIT spectra. (b) Simulated contour plot of power broadening and splitting in the CL configuration. For both (a) and (b) $\Omega_{p} = 2\pi\times 2.4$ MHz and $\Omega_{R} = 2\pi\times1.2$ MHz.}
    \label{fig:contour_plot}
\end{figure}

 The contour plot of Figure~\ref{fig:contour_plot} shows the CL configuration measured EIT spectra along with simulated results over a range of dressing Rabi frequencies. Due to a larger volume and thus larger optical depth in the CL configuration (relative to the DF configuration), we can resolve EIT spectra at lower dressing Rabi frequencies where power broadening is further reduced. Clear splitting is observed above dressing Rabi frequencies of $\approx 2\pi\times10$ MHz in both experiment and simulation. The inset in Fig.~\ref{fig:contour_plot} (a) shows the EIT spectra for both experiment and simulation at the lowest measured dressing Rabi frequency of $2\pi\times 2$ MHz. The experimentally measured FWHM is $3.6(3)$ MHz, which is in reasonable agreement with our simulated prediction of $3.1$ MHz. The data is fit to a Voigt profile where the widths of the Gaussian and Lorentzian contributions are treated as free parameters with the FWHM being computed numerically. Further the asymmetric shape is accounted for by allowing these widths to vary on each side of the peak. The asymmetry in the experimental spectra is likely due to a slight detuning of the dressing laser. We use the same dephasing terms for the simulation results presented in Fig. \ref{fig:contour_plot} (b) and Fig. \ref{fig:angular_results}. Note that without these dephasing parameters, the decrease in peak amplitude with increasing $\Omega_{d}$ is not captured.

The narrowest EIT spectra occur at the lowest probe and dressing Rabi frequencies, as seen in Fig.~\ref{fig:power_broadening} (a). Even at our lowest powers the EIT linewidths are power broadened Lorentzian profiles. Increasing these Rabi frequencies results in additional power broadening and can produce asymmetric features in the spectra. Fig.~\ref{fig:power_broadening} (a) gives the FWHMs from the DF configuration as a function of dressing Rabi frequency for various probe Rabi frequencies. Fig.~\ref{fig:power_broadening} (b) gives the results for the narrowest DF linewidth observed for $\Omega_{p} = 2\pi\times2.2$ MHz, $\Omega_{d} = 2\pi\times21.4$ MHz, and $\Omega_{R} = 2\pi\times1.2$ MHz along with the CL spectra for comparable Rabi frequencies. There is a clear distinction in that the CL spectra shows a peak splitting, whereas in DF the spectra of the various velocity classes combine to a single peak. Fitting the CL spectral feature to a double Voigt, the FWHM of the individual peaks is 4.36(6) MHz. The asymmetry of the peak heights is due to a slight detuning, of order a few MHz, of the dressing laser. In contrast, the DF spectra does not exhibit splitting and shows a 3.7 times reduction in the FWHM at 1.18(8) MHz.

\begin{figure}[t!]
    \centering
    \includegraphics[width=0.48\textwidth]{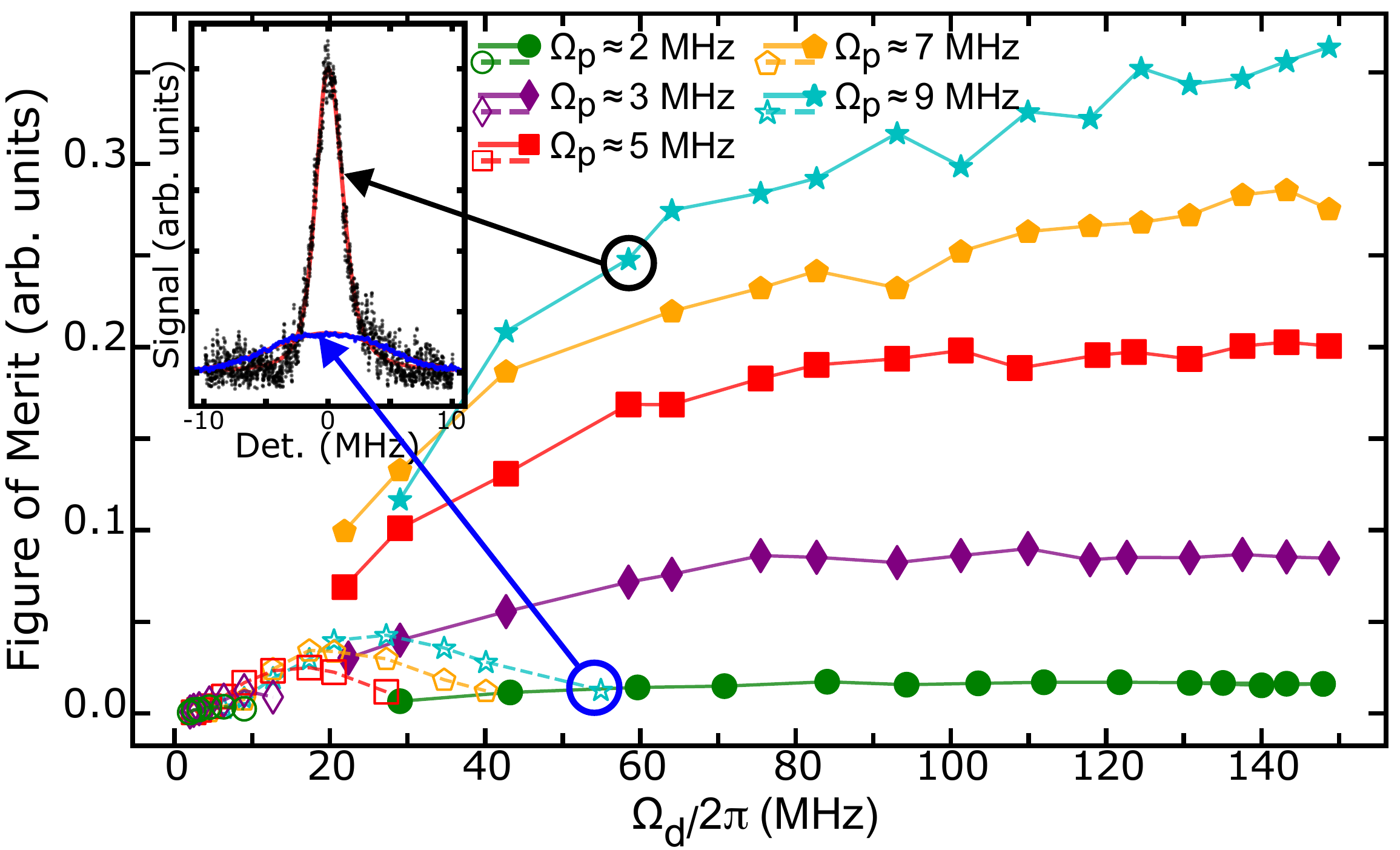}
    \caption{Comparison of DF and CL figure of merit, defined as per unit volume ratio of signal amplitude-to-FWHM. Figure of merit is plotted versus $\Omega_{d}$ for various $\Omega_{p}$ (represented with symbols and colors) and $\Omega_{R}$ held at constant 1.2 MHz. Multiply legend labels by $2\pi$ to retrieve the Rabi frequencies in proper units. Solid line and closed symbol are DF, while dashed-line and open symbol are CL. We neglect CL spectra when they split to two peaks as $\Omega_{d}$ increases (see Fig. \ref{fig:contour_plot}). Inset shows volume-normalized DF and CL spectrum data. The DF system has a greater figure of merit than the CL (volume normalized) for a wide range of Rabi frequencies.}
\end{figure}

As dominant line broadening mechanisms, such as Doppler effects and power broadening, are reduced, contributions from other mechanisms must be considered. Next relevant mechanisms for our experiment include: shifts of magnetic sublevels from background magnetic fields, laser noise, stray electric fields, and atom-atom interactions. The background magnetic field in the vicinity of the vapor cell is primarily from Earth's  field and is on the order of 0.5 Gauss. Laser linewidths on the relevant timescales are all less than 200 kHz FWHM, but simulations suggest that such laser noise will begin to play a role in our measured spectra. For example, simulations show that with negligible laser noise the EIT linewidth as a function of $\Omega_{d}$ asymptotes to a constant value, whereas including white laser noise causes conitinued rise of EIT linewidth, like we observe in Fig. \ref{fig:power_broadening}(a). Finally, as the above sources are reduced, stray fields, for example from surface charge, and atom-atom interactions (including Rydberg-Rydberg interactions) can contribute to spectral broadening. Comprehensive modeling and measurement of these effects is beyond the scope of this manuscript, and will be the focus of subsequent work.

\begin{figure}[t!]
    \centering
    \includegraphics[width=0.5\textwidth]{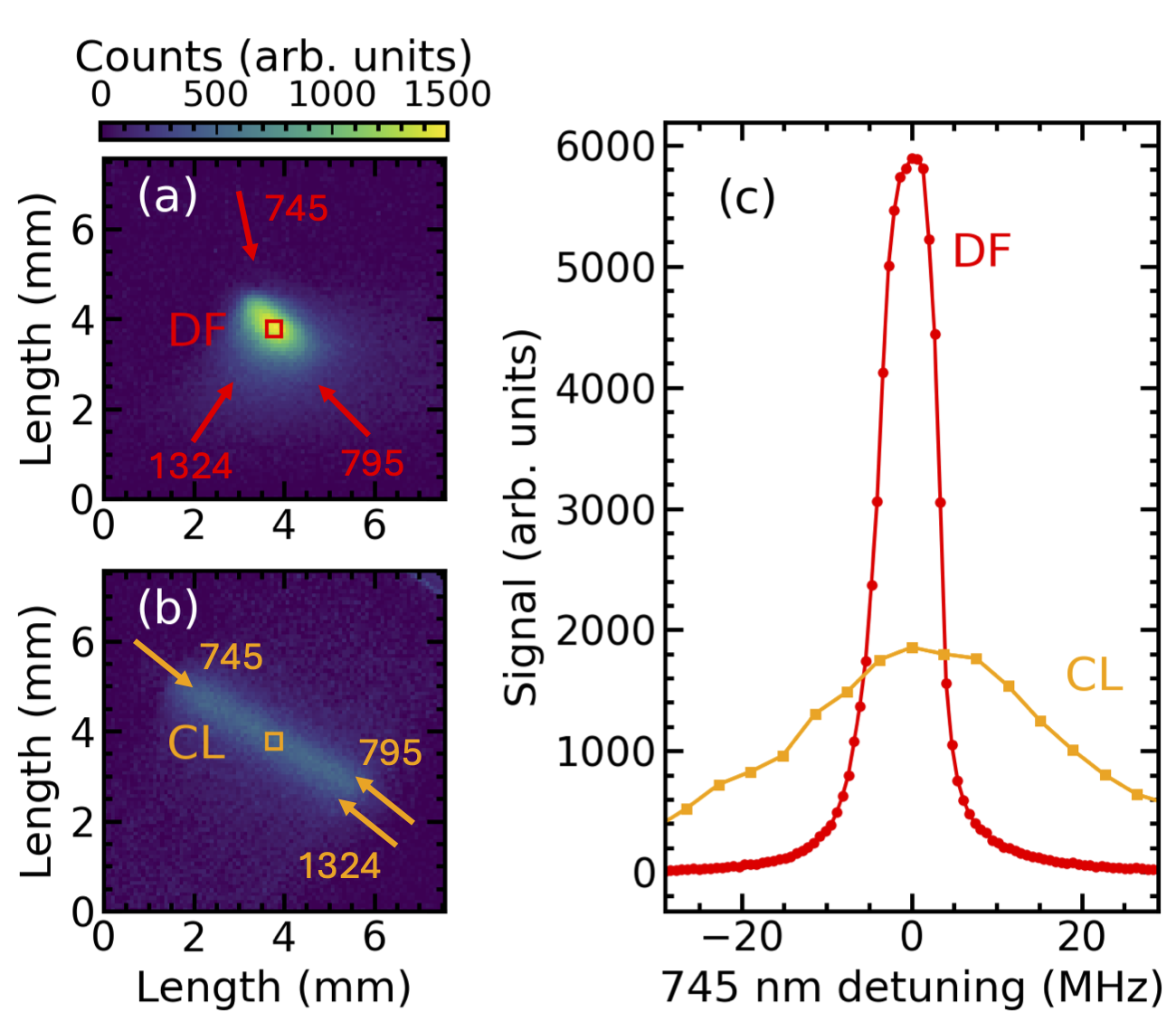}
    \caption{Fluorescence artificial-color image results for DF (a) and CL (b) configurations. Counts are integrated over the frequency scan range of the Rydberg  laser. The regions of highest signal within a $0.34\times0.34$ mm region are indicated by the red and orange boxes. The fluorescence spectra as a function of Rydberg laser detuning for those regions are shown in (c). The data was taken with the Rabi frequencies $\Omega_{p}=2\pi\times 19.2$ MHz, $\Omega_{d}=2\pi\times 136.7$ MHz, and $\Omega_{R}=2\pi\times1.2$ MHz. }
    \label{fig:image_comparison}
\end{figure}

To consider the potential sensing capabilities of the CL vs DF systems, we define a figure of merit that is the ratio of the resonance amplitude to FWHM normalized to a unit volume (directly related to sensitivity per unit volume), and the results are shown in Fig. \ref{fig:discriminator}. This figure of merit is particularly relevant for sensing applications where fine spatial resolution is important or size restrictions limit the sensing volume. The scale factor to normalize the volumes between the two geometries in our case is $V_{DF}/V_{CL}\approx0.13$. The inset in Fig. \ref{fig:discriminator} shows an example comparison between scaled DF and CL spectra at $\Omega_d/2\pi\approx60$ MHz and $\Omega_p/2\pi = 9$ MHz. Due to the lower density of Rydberg atoms, the CL spectra peaks are weaker, and because of Doppler effects they are also broader indicating that DF configuration offers advantages in sensing especially for high spatial resolution measurements. The trends in Fig. \ref{fig:discriminator} show that for both systems, there is an initial increase in the signal to FWHM ratio with an increase of either probe or dressing laser power. The CL data peaks and then decrease due to unresolved splitting of intermediate states (e.g. Fig. \ref{fig:contour_plot}). We are unable to measure the DF signals at the smallest $\Omega_d$ values, however we see a similar trend as in CL.

To directly compare the relative Rydberg densities in the CL and DF geometries we also performed florescence measurements. We collect blue fluorescence with a bandpass filter from $\approx$ 462 to 512 nm that is emitted via numerous decay pathways from the Rydberg state as illustrated in Fig.~\ref{fig:system} (d). The strongest decay channels which emit blue fluorescence are decays to $29D_{3/2}$ and $30D_{3/2}$ via blackbody radiation with estimated transition rates of $\approx 2\pi \times1$ kHz and $\approx 2\pi\times0.6$ kHz \cite{Weatherill_2017_ARC}. Given these slow coupling rates along with the fast thermal velocity of the atoms and narrow laser beam waist, an atom on average will emit no more than one approximately 480 nm photon while traversing through the beams. Hence, measuring the blue fluorescence provides a reasonable estimate for determining the relative Rydberg densities between the two configurations.

Figs.~\ref{fig:image_comparison} (a) and (b) show the detuning frequency integrated fluorescence results in the DF and CL configurations respectively. Here, the fluorescence signal is integrated over all Rydberg laser detunings. In Figs.~\ref{fig:image_comparison} (c) and (d) are the fluorescence spectra in the DF and CL configurations respectively as a function of Rydberg laser detuning. The signal is integrated spatially over the red and orange boxes shown in Fig.~\ref{fig:image_comparison} (a) and (b). The peak, on resonance Rydberg density is $\approx 3$ times higher in the DF configuration.

\section{Conclusion}
We have presented results on warm vapor Rydberg coherent spectroscopy comparing the common collinear (CL) laser geometry to a Doppler-free (DF) star configuration. This represents, to the best of our knowledge, the first report of a narrowed Rydberg EIT linewidth resulting from Doppler cancellation in a ``star'' configuration. 
This approach allows for a nearly arbitrary choice of desired Rydberg state excitation, and convenient choice of laser wavelengths, which could be important for making Rydberg sensors more widely applicable. We show that the DF system results in significantly narrower linewidths, and a three times increase in Rydberg atom density. We also show important features such as increasing Rabi frequencies without spectral splitting. Taken together, the DF method results in an increased signal slope per unit volume, which can be a critical figure of merit for sensing applications where fine spatial resolution is desired ~\cite{Meyer_2024_polarimetry}. We note that special consideration must be taken in the orientation and geometry of the vapor cell to avoid deleterious reflections. While we have focused on removing the Doppler component to EIT broadening using a star configuration, possibilities remain for further narrowing of the EIT linewidths. This includes reducing residual external magnetic and electric fields, transit broadening and laser noise which will be part of future work. 

\begin{acknowledgments}
We gratefully acknowledge helpful discussion and early contributions from Donald Fahey.
\end{acknowledgments}

\section{References}
\bibliographystyle{apsrev4-2}
\bibliography{refs.bib}

\end{document}